\documentstyle[graphicx,aps,multicol]{revtex}
\draft
 
\begin{document}

\title{Level densities and $\gamma$-ray strength functions 
in $^{170, 171, 172}$Yb}

\author{U.~Agvaanluvsan,$^{1}$\footnote{Electronic address: agvaanluvsan1@llnl.gov}
A.~Schiller,$^{2}$ J.~A.~Becker,$^2$ L.~A.~Bernstein,$^2$ P.~E.~Garrett,$^2$ 
M.~Guttormsen,$^3$ G.~E.~Mitchell,$^{1,4}$ J.~Rekstad,$^3$ S.~Siem,$^3$ A.~Voinov$^5$, 
and W.~Younes$^2$}
\address{$^1$Department of Physics, North Carolina State University, Raleigh, NC 27695,
USA}
\address{$^2$Lawrence Livermore National Laboratory, L-414, 7000 East Avenue,
Livermore, CA 94551, USA}
\address{$^3$Department of Physics, University of Oslo, N-0316 Oslo, Norway}
\address{$^4$Triangle Universities Nuclear Laboratory, Durham, NC 27708, USA}
\address{$^5$Department of Physics, Ohio University, Athens, OH 45701, USA}

\maketitle

\begin{abstract}
Level densities and radiative strength functions in
$^{171}$Yb and $^{170}$Yb nuclei have been measured
using the $^{171}$Yb($^3$He,$^3$He$^\prime\gamma$)$^{171}$Yb and
$^{171}$Yb($^3$He,$\alpha\gamma$)$^{170}$Yb
reactions.
New data on $^{171}$Yb are compared to a previous measurement 
for $^{171}$Yb  from the 
$^{172}$Yb($^3$He,$\alpha\gamma$)$^{171}$Yb reaction.
Systematics of level densities and radiative strength functions in 
$^{170,171,172}$Yb are established. 
The entropy excess in $^{171}$Yb relative to the even-even nuclei $^{170,172}$Yb 
due to the unpaired neutron quasiparticle is found to be approximately 2$k_B$.  
Results for the radiative strength function from the two reactions lead
to consistent parameters characterizing the ``pygmy'' resonances.
Pygmy resonances in the $^{170,172}$Yb populated by the 
($^3$He,$\alpha$) reaction appear to be split into two components
for both of which a complete set of resonance parameters are obtained.

\end{abstract}  

 
\begin{multicols}{2}
 
\section{Introduction}   
Nuclear level densities and radiative strength functions
are important inputs for calculations of
nuclear reaction cross sections. In addition to their value in  practical
applications,
 these average
quantities may shed light on the understanding of such fundamental 
issues  as the transition  
from the discrete (low excitation) 
to statistical (high excitation) regime. 
At low excitation energy, the level density is obtained directly 
by counting low-lying levels. 
However, at increasing excitation energy,
the level density becomes large and individual
levels are often not resolved in experiments\cite{Gil65}.  Nuclear resonances 
at or above the nucleon binding energy provide another source 
of level density data \cite{Egi88}. Between these two
excitation energy regions, relatively
little is known about nuclear level densities. 
The present paper focuses on this intermediate region.

A major part of the information on radiative strength functions comes from
 photoabsorption cross section measurements \cite{Die88}.
High energy $\gamma$ transitions  ($E_\gamma\sim 10-15$ MeV) 
are dominated by the giant electric dipole resonance (GEDR).
Although the electric dipole transition strengths are well
studied in the vicinity of the GEDR, the behavior of low energy 
$\gamma$ rays 
is less well understood \cite{Kop90}.  This is particularly true for radiative 
transitions between highly excited states.  Experimental data on
the $M1$ strength function are much scarcer 
 than for the $E1$ strength function.
In these regions, an experimental technique recently developed 
by the Oslo Cyclotron Group provides valuable data. 
This method allows one to determine level densities
and radiative strength functions simultaneously \cite{Sch00a,Hen95}
from the primary $\gamma$-ray spectra. 
The advantage of this method is that it provides data
on nuclear level densities and radiative strength functions
in regions where there is little information and data is difficult to obtain.
However, the level density and radiative strength function are coupled, 
since the $\gamma$ decay input to the technique depends on both quantities. 
A shortcoming of the method is that the absolute level density
and radiative strength function need to be normalized 
using the low-lying discrete states, neutron
resonance spacings, and average total radiative widths of neutron resonances. 
Thus the primary new contribution is the energy dependence of the
level density and the radiative strength function.  
This method is commonly referred to as the Oslo method.
It has been shown to work well in heavy-mass nuclei and
has been extended to other mass regions as well \cite{Tav02,Gut03c}.
The present paper reports new results from a
$^{171}$Yb + $^3$He experiment. The Oslo method and the
experimental set-up are briefly discussed, followed by a brief
description of level densities and radiative strength functions.
The results for $^{171}$Yb obtained from two different reactions, 
$^{171}$Yb($^3$He,$^3$He$^\prime$)$^{171}$Yb 
and $^{172}$Yb($^3$He,$\alpha$)$^{171}$Yb,
are compared. The similar comparison for $^{172}$Yb, previously reported,
is repeated for the sake of completeness.

\section{Experimental methods} 

The experiment was conducted at the Oslo Cyclotron Laboratory (OCL)
using a 45-MeV $^3$He beam. The self-supporting targets of
$^{171,172,173}$Yb enriched to $\sim$ 95 \% 
had a thickness of $\sim$ 2 mg cm$^{-2}$. Five reactions
studied in this paper are: \\
1) $^{171}$Yb($^3$He,$\alpha$)$^{170}$Yb (new) \\      
2) $^{171}$Yb($^3$He,$^3$He$^{\prime}$)$^{171}$Yb (new)\\
3) $^{172}$Yb($^3$He,$\alpha$)$^{171}$Yb (reported previously in \cite{Sch01,Voi01}) \\
4) $^{172}$Yb($^3$He,$^3$He$^{\prime}$)$^{172}$Yb (reported previously in 
\cite{Sch00})\\ 
5) $^{173}$Yb($^3$He,$\alpha$)$^{172}$Yb (reported previously in \cite{Sch01,Voi01}).

Particle-$\gamma$ coincidences for $^{170,171,172}$Yb were detected using the
CACTUS multidetector array \cite{Gut90}. The charged particles were measured with eight
Si particle telescopes placed at 45$^\circ$ with respect to the beam
direction. Each telescope consists of a front Si $\Delta E$ detector and a back
Si(Li) $E$ detector with thicknesses 140 and 3000 $\mu$m, respectively.
An array of 28 collimated
NaI $\gamma$-ray detectors with a total coverage
of $\sim$ 15\% of 4$\pi$ surrounds  the target. In addition one to three Ge detectors
were used to estimate the spin distribution and determine the
selectivity of the
reaction. The typical spin range is expected to be
$I\sim 2-6 \hbar$. Experiments run for $\sim$ 2 weeks with typical beam currents of $\sim$ 1.5 nA. 
 
The data analysis consists of three main steps. The first step is 
to prepare the particle-$\gamma$ coincidence 
matrix. For each particle energy bin total cascade $\gamma$-ray 
spectra are obtained from the coincidence
measurement. The particle energy is transformed
to  excitation energy using the reaction kinematics.
Then each row of the coincidence matrix corresponds to a certain excitation 
energy $E_x$ in the residual nucleus, while each column corresponds to a certain
$\gamma$-ray energy $E_\gamma$. The second step is the unfolding.
The $\gamma$-ray spectrum is unfolded using the known response
function of the CACTUS array \cite{Gut96}. 
The $\gamma$-ray spectrum containing  only
the first $\gamma$ rays emitted from a given  excitation energy
is called the first-generation spectrum and denoted by $P$. The matrix which consists 
of first generation spectra is obtained in the third step
for each excitation energy bin using the
subtraction procedure described in Ref.\ \cite{Gut87}. 
The key assumption of this method is that the $\gamma$ decay from any 
excitation energy bin is independent of the 
method of formation, 
either  directly by the nuclear reaction or by $\gamma$ decay from higher
lying states following the initial reaction. This assumption is
automatically fulfilled when the same states are  populated equally
via the two processes, since
$\gamma$ branching ratios are properties of levels. Even if different
states are  populated, 
 the assumption is still valid for statistical $\gamma$ decay, which 
depends only on the $\gamma$-ray  energy and the number of available 
final states.
These assumptions have been investigated extensively over the years by 
the Oslo group and shown to work reasonably well \cite{Hen95}.  
The entries of the first generation matrix $P$
are the probabilities $P(E_x,E_\gamma)$ that
a $\gamma$-ray of energy $E_\gamma$ is emitted from
excitation energy $E_x$.

The probability of $\gamma$ decay
is proportional to the product of the $\gamma$-ray strength
(i.e., the radiative transmission coefficient
${\cal T} (E_\gamma)$) and
the level density $\rho(E_x-E_{\gamma})$ at the final energy $E_x-E_{\gamma}$
\begin{equation}
\label{probab}
P(E_x,E_\gamma)\propto {\cal T}(E_\gamma)\rho(E_x-E_\gamma).
\end{equation}
This factorization is
the generalized form of the Brink-Axel hypothesis \cite{Bri55,Axe62} 
which states that the GEDR and any other excitation modes built on 
an excited state have the same properties as those built 
on the ground state. 
In other words, the radiative transition strength is independent of
the excitation energy.  
There is evidence that the width of the giant-dipole resonance
increases with increasing nuclear temperature of the state 
upon which it is built, and thus with its 
excitation energy $E_f=E_x-E_\gamma$ \cite{Kad83,Ger98}.  
The temperature that corresponds to the excitation energy
region covered in this work is rather low and changes slowly 
with excitation energy ($T\sim \sqrt{E_f}$). In this work we assume 
constant temperature and  that the radiative strength function does not 
depend on the excitation energy in the energy
interval under consideration.

In the Oslo method the functions $\rho$ and ${\cal T}$ are determined
by an iterative procedure 
\cite{Sch00a}. The goal of the iteration
is to determine these two functions at $\sim N$ energy values each;
the product of the two functions is known at $\sim N^2/2$ data points.
The globalized fitting to the data points determines
the functional form for $\rho$ and 
${\cal T}$ \cite{Sch00a}. However, it can be
shown that the entries of matrix $P$ in Eq.\ (\ref{probab}) given 
by the product of $\rho$ and ${\cal T}$ are invariant under
the transformation \cite{Sch00a}

\begin{eqnarray}
\tilde{\rho}(E_x-E_\gamma)&=&A\exp[\alpha(E_x-E_\gamma)]\,\rho(E_x-E_\gamma),
\label{eq:array1}\\
\tilde{\cal T}(E_\gamma)&=&B\exp(\alpha E_\gamma) {\cal T}(E_\gamma).
\label{eq:array2}
\end{eqnarray}

Thus, in a final step, the transformation parameters
$A$, $B$, and $\alpha$, which correspond to the physical solution 
have to be determined.
The level density $\rho$ is determined from the nuclear ground state
up to $\sim B_n-$1 MeV,  where $B_n$ is the neutron binding energy.
The coefficients $A$ and $\alpha$ are determined from 
the normalization of the level density to data from 
the discrete levels and the neutron resonance spacings.
The radiative transmission coefficient ${\cal T}$ is obtained from
$E_\gamma\approx$ 1 MeV to around $B_n$. The remaining constant 
$B$ is determined from the normalization of the transmission coefficient
to data from the total radiative widths of neutron resonances.
The details of the normalization and consideration  of 
the experimental results are discussed in the following sections. 

\section{Level densities} 

\begin{figure}
\includegraphics[totalheight=21cm,angle=0,bb=0 80 350 730]{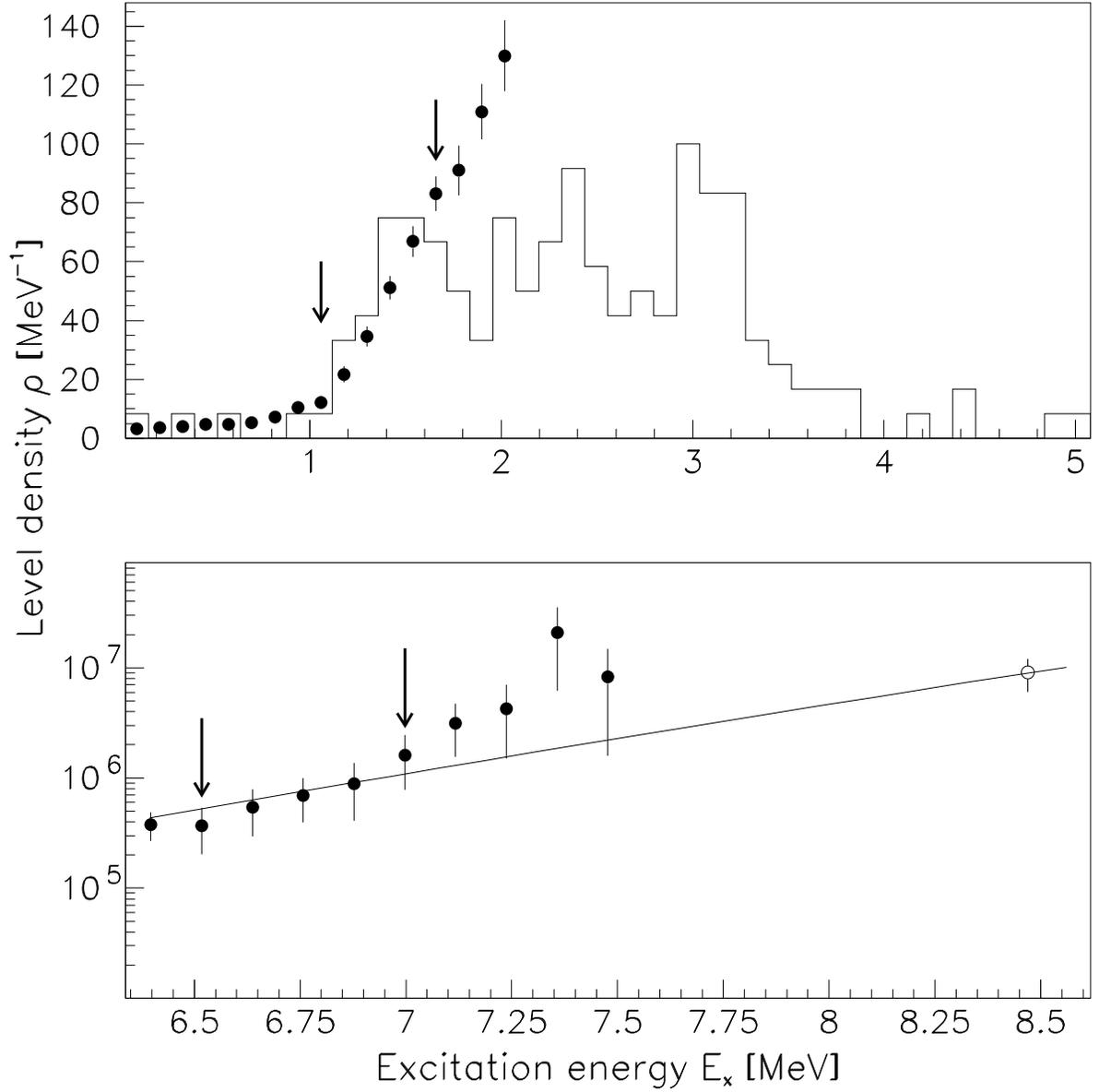}
\caption{Normalization procedure for 
 the experimental  $^{170}$Yb level density. 
The experimental data from this paper are represented by  full circles. 
In both panels, arrows enclose the data points used for normalization. 
The data were fit to discrete levels (shown as histograms in the upper
panel)
 and to the level density calculated using the resonance spacings 
(shown as an open circle in the lower panel).
The Fermi gas level density, Eq.\ \ref{rhoEq}, (line) was employed to connect the regions
where data were available.}
\label{counting170yb}
\end{figure}

The level density obtained from the present experiment covers 
the excitation energy from the ground state up to $\sim B_n-$1 MeV\@.
However, as described in the previous section, the level density 
must  be normalized.
Figure \ref{counting170yb} illustrates  the normalization procedure.
The filled circles are our experimental data points for the reaction
$^{171}$Yb$(^3$He,$\alpha)^{170}$Yb.
In the upper panel, the level density at low excitation energy
determined from the present experimental data is 
compared to the histogram calculated using 
discrete levels listed in the Table of Isotopes \cite{Fir96}.
The agreement is good up to $E_x\sim$ 1.6~MeV\@.
Above this energy  the two results
differ because there is limited information on discrete levels at higher
excitation energy. Thus, the present experiment provides new results for
the average level density above
$E_x=$ 1.6 MeV\@. Comparison at low excitation energy is used to fix the
absolute value of the data at the low-energy end.
In the lower panel, the normalization with respect to the neutron 
resonance spacing data is shown. 
The level density at $B_n$ is determined using the 
neutron resonance spacing data \cite{RIPL}.
Since our data only extend to $\sim$ 1~MeV below $B_n$,
an  interpolation is required  
between the present experimental data and $\rho$ evaluated
at $B_n$. The back-shifted Fermi gas level density
parameterized by von Egidy {\sl et al.} \cite{Egi88}, 
\begin{equation}
\label{rhoEq}
\rho(E_x)=\eta\frac{\mathrm{exp}(2\sqrt{aU})}{12\sqrt{2}a^{1/4}U^{5/4}\sigma_I},
\end{equation}
is employed for the interpolation.
The back-shifted excitation energy is given by $U=E_x-C_1-\Delta$ where
$C_1=-6.6A^{-0.32}$~MeV, $a=0.21\ A^{0.87}$~MeV$^{-1}$ and the pairing parameter 
$\Delta$ is estimated following the prescription by 
Dobaczewski {\sl et al.} \cite{Dob01}.     
For $^{172}$Yb, a slightly different normalization was used.
The level density for $^{170}$Yb is about an order of magnitude smaller
than the level density for $^{171}$Yb. 

We report new results for the level density and radiative strength function 
obtained from reactions on the target nucleus $^{171}$Yb. 
For $^{171}$Yb, the level density and radiative
strength function were obtained by two different reactions,
one from the $^{171}$Yb$(^3$He,$^3$He$^\prime\gamma)^{171}$Yb reaction and 
another 
from the $^{172}$Yb$(^3$He,$\alpha\gamma)^{171}$Yb reaction.
The results from the 
$^{172}$Yb($^3$He,$\alpha\gamma$)$^{171}$Yb experiment were reported
previously in \cite{Sch01,Voi01}.
Similarly, the level density in $^{172}$Yb is obtained by two
reactions. 

\begin{figure}
\includegraphics[totalheight=20cm,angle=0,bb=0 0 350 830]{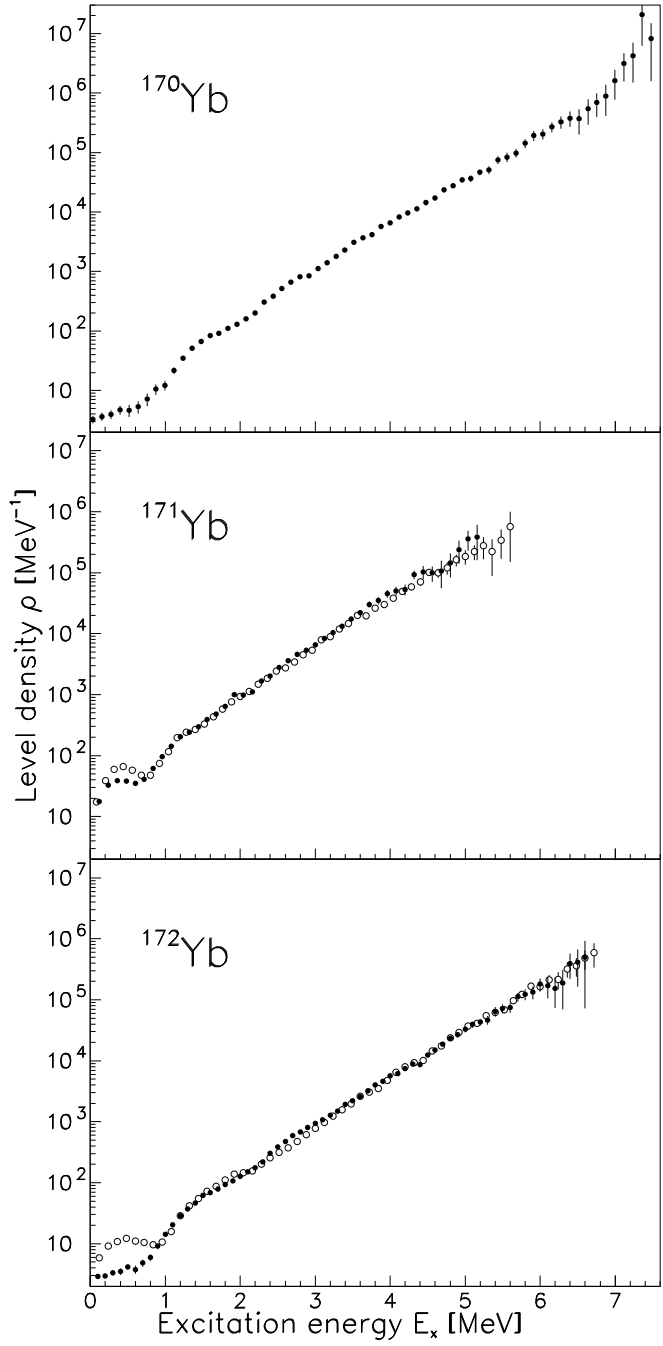}
\caption{Deduced level densities for $^{170,171,172}$Yb. The full and open circles correspond to data obtained from the ($^3$He,$\alpha$) and ($^3$He,$^3$He$^{\prime}$) reactions, respectively.}
\label{rhoall}
\end{figure}

Figure \ref{rhoall} shows the level densities of $^{170,171,172}$Yb
from the ground state
up to $\sim B_n-$ 1 MeV\@. Data points shown in full circles are from
the ($^3$He,$\alpha\gamma$) reaction
and in open circles from the ($^3$He,$^3$He$^\prime\gamma$) reaction.
The effect which yields an overestimated level density at low excitation 
$\sim$ 0.5~MeV from $(^3$He,$^3$He$^\prime)$ reaction data is due to a 
disporportional population of states with relatively large 
transition matrix elements to the ground state rotational band.
More details on this effect are given in Ref.\ \cite{Sch00b}. 
Except for this effect, 
the agreement between level densities for the same
nucleus obtained via two different reactions is excellent.   
 
The level density is closely connected to the entropy $S$ of
the system at a given excitation energy $E_x$. This 
opens the possibility of investigating certain thermodynamic
properties in the atomic nucleus.
The entropy is given by
\begin{equation}
S(E_x)=k_{\rm B}\ln \Omega(E_x).
\end{equation}
The Boltzmann's constant $k_B$ is set to unity from here on.
The multiplicity $\Omega$ is directly proportional to the level
density:  $\Omega(E_x)= \rho(E_x)/\rho_0$. The ground states of even-even
nuclei represents a well-ordered system with no thermal excitations
and are characterized by zero entropy and temperature.
Therefore the normalization denominator is set
to $\rho_0=3$ MeV$^{-1}$ to obtain $S=\ln \Omega \sim 0$ in the ground
state band region. This ensures that the ground band properties
fulfill the third law of thermodynamics with $S(T\rightarrow 0) = 0$.
The extracted $\rho_0$ is also used for the odd-mass neighboring nuclei.
 
\begin{figure}
\includegraphics[totalheight=20cm,angle=0,bb=0 0 350 830]{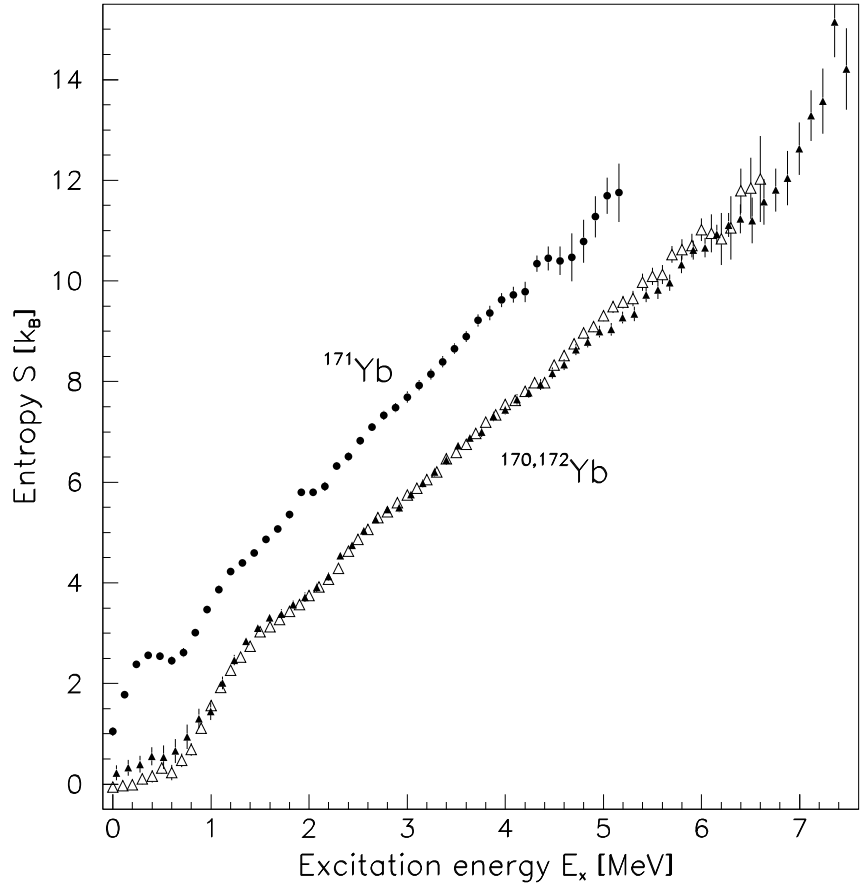}
\caption{Deduced entropies for $^{170,171,172}$Yb. All data are from 
the ($^3$He,$\alpha$) reaction. The full circles correspond to $^{171}$Yb. 
The full and open triangles correspond to $^{170}$Yb and $^{172}$Yb, 
respectively.}
\label{ybentr}
\end{figure}

Figure \ref{ybentr} shows the entropies $S$ of $^{170,171,172}$Yb obtained
from the ($^3$He,$\alpha\gamma$) reaction. Several properties derived from this
figure are actually connected to the fact that these mid-shell
rare earth nuclei have similar nuclear structure and global properties,
such as nuclear deformation.
 
The entropies (or level densities)
of $^{170}$Yb and $^{172}$Yb follow each other closely
as a function of excitation energy. In particular, in the excitation energy
region from the ground state up to 2 MeV, 
$S(E_x)$ shows very similar shapes. We interpret
the strong increase around 1.5 MeV of excitation energy as
the breaking of the first Cooper pair. The next increase,
which is much more smeared out, terminates near 2.5 MeV and
reveals the beginning of the four quasiparticle regime.
Above 2.5 MeV the entropy increases linearly \cite{Gut03a}.

In addition, all three entropy curves are 
parallel for excitation energies above $E_x\sim 2.5$ MeV\@. In the
microcanonical ensemble, the slope of $S(E_x)$ is connected to
the temperature by
\begin{equation}
T=(dS/dE_x)^{-1}_V.
\end{equation}
A constant temperature least-square fit in the $E_x=2.5 - 5.5$ MeV region of
$^{170,171,172}$Yb gives $T= 0.62(3), 0.52(4)$ and $0.58(3)$ MeV, respectively.
These temperatures are interpreted as the critical temperatures $T_c$ for the 
breaking of nucleon pairs.
 
It is interesting to note that the entropy of $^{171}$Yb shows a
strongly increasing behavior that also terminates at $E_x \sim 1.5$ MeV
revealing the first breaking of Cooper pairs in the underlying
even-even core. Here the odd valence nucleon behaves as a passive
spectator, however,  the increase in $S(E_x)$
appears at slightly lower excitation energies than for the even systems.
This behavior is attributed to the reduced pairing gap $\Delta$ resulting
from the Pauli blocking  by the valence neutron in the odd system.
 
\begin{figure}
\includegraphics[totalheight=20cm,angle=0,bb=0 0 350 830]{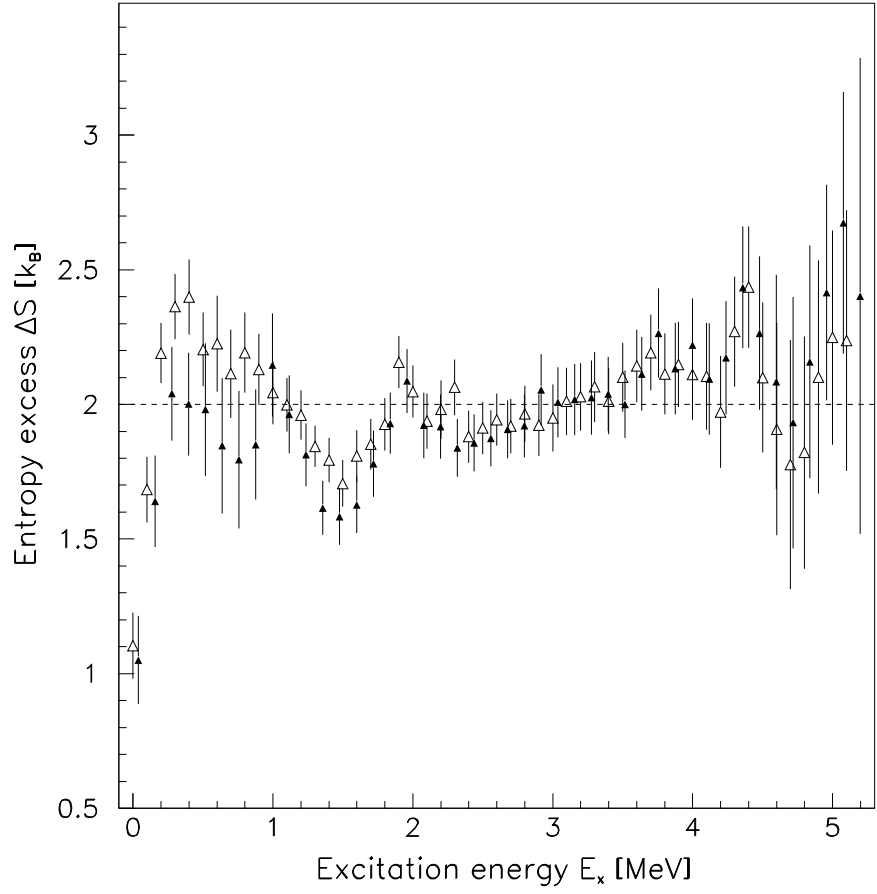}
\caption{Deduced entropy excess for $^{171}$Yb. The full and open triangles correspond 
to entropy excesses for the single particle and hole, respectively, calculated 
using Eqs.\ (\ref{entrex1}) and (\ref{entrex2}). 
The entropy excess $\Delta S=$ 2$k_B$ is shown by the dashed line.}
\label{ybdsn}
\end{figure}

The entropy carried by the valence neutron particle (or hole) can
be estimated assuming that the entropy is an extensive (additive) 
quantity \cite{Gut01}.
Figure \ref{ybdsn} shows the observed single particle and hole
entropies defined by
\begin{eqnarray}
\label{entrex1}
\Delta S({\rm particle})&=&S(^{171}{\rm Yb})- S(^{170}{\rm Yb}),\\
\label{entrex2}
\Delta S({\rm hole})    &=&S(^{171}{\rm Yb})- S(^{172}{\rm Yb}),
\end{eqnarray}
respectively. The single particle (or hole) carries
about $\Delta S=2$. Deviations from
this estimate appear at low excitation energies due to the lower pairing
gap in the odd-mass  system. At higher energies the slightly lower
critical temperature in the odd-mass system is responsible for the
increasing entropy difference as function of excitation energy.
These two qualitative explanations are connected.
The pairing gap $\Delta$, critical temperature $T_c$ and single
particle (hole) entropy $\Delta S$ are related \cite{Gut01} by
\begin{equation}
T_c=\frac{1}{\Delta S}\Delta,
\end{equation}
for constant $\Delta S$. This is consistent with the present
observations. Thus, it is reasonable to expect that if
both $\Delta$ and $T_c$ were equal in the systems compared,
the ${\Delta}S$ curve would be flatter as function
of excitation energy.
 
The thermodynamical properties can also be studied using the
canonical ensemble. Recently \cite{Gut03b} this was performed for the
$^{160,161,162}$Dy isotopes that behave very much in the
same way as the present Yb isotopes.

\section{Gamma-ray strength functions}
  
\begin{figure}
\includegraphics[totalheight=20cm,angle=0,bb=0 10 350 830]{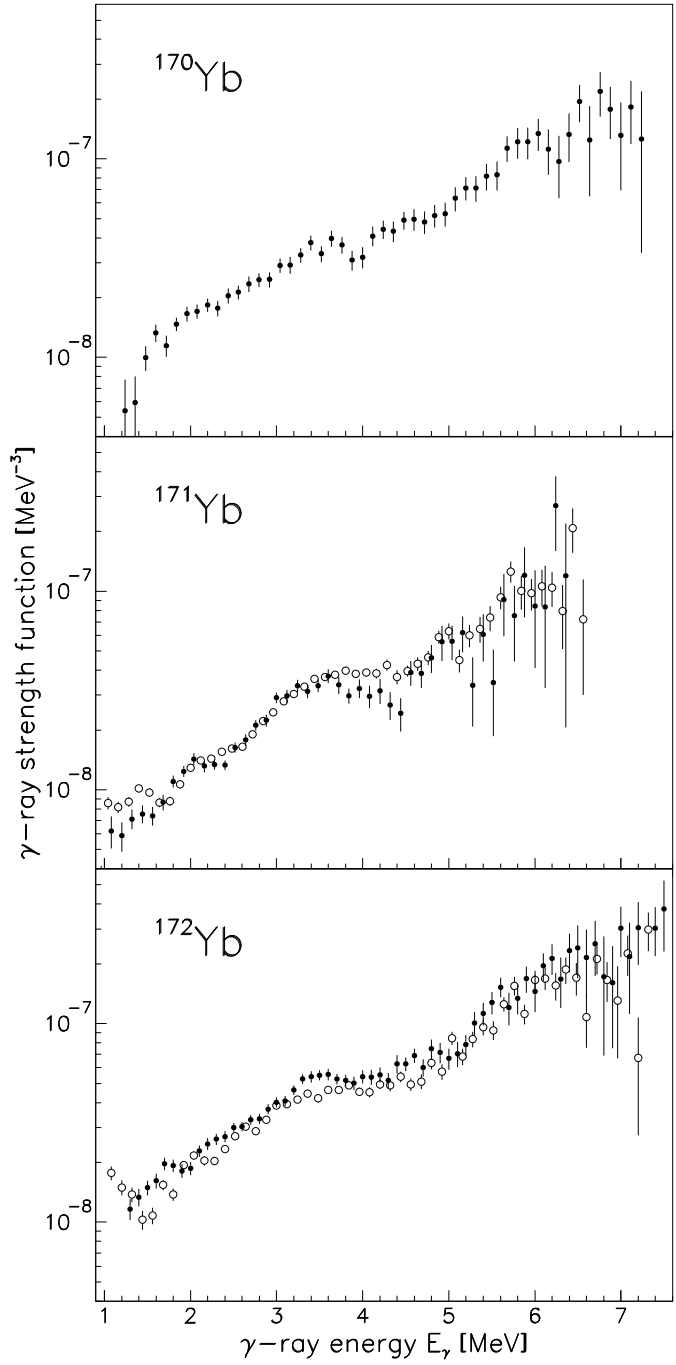}
\caption{Radiative strength functions for $^{170,171,172}$Yb. The full and open circles correspond to data obtained from the ($^3$He,$\alpha$) and ($^3$He,$^3$He$^{\prime}$) reactions, respectively.}
\label{strall}
\end{figure}

The $\gamma$-ray transmission coefficient ${\cal T}(E_{\gamma})$
in Eq.\ (\ref{probab}) is expressed as a sum of all
the $\gamma$-ray strength functions $f_{XL}$ of multipolarities $XL$
\begin{equation}
{\cal T}(E_\gamma)=2\pi\sum_{XL}E_{\gamma}^{2L+1}f_{XL}(E_\gamma).
\end{equation}
The radiative transmission coefficient ${\cal T}$ obtained from the present work
is unnormalized. As shown in Eqs.\ (\ref{eq:array1},\ref{eq:array2}) and in the
previous section, two of the three normalization coefficients are
obtained from the level density. The remaining constant $B$ in 
Eq.\ (\ref{eq:array2}) is determined using information from neutron
resonance decay.  
The average radiative width of neutron resonances $\langle\Gamma_\gamma\rangle$
at the neutron binding energy is related to ${\cal T}(E_{\gamma})$
\begin{equation}
\label{avgGamma}
\langle\Gamma_\gamma\rangle=\frac{B}{4\pi\rho(B_n,J_i^\pi)}
\sum_{J_f^\pi}\int_0^{B_n} {\mathrm{d}}E_{\gamma}{\cal T}(E_{\gamma})
\rho(B_n-E_{\gamma},J_f^\pi),
\end{equation}
where $D_i=1/\rho(B_n,J_i^\pi)$ is the average spacing of 
$s$-wave neutron resonances, and the sum extends over all possible
final state spins and parities, and matching multipole contribution to
${\cal T}(E_\gamma)$. The level density is assumed to have 
the standard energy and spin dependent parts
\begin{equation}
\rho(E_x,J)=\rho(E_x)\frac{2J+1}{2\sigma^2}e^{-(J+1/2)^2/2\sigma^2},
\end{equation} 
where $\sigma$ is the spin cut-off parameter, and we assume equal number of
positive and negative parity states. The spin cut-off parameter
is calculated as a function of excitation energy by a linearization of the 
usual $\sigma\sim U^{1/4}$ around $B_n$  
\begin{equation}
\sigma=\sigma_0 \left( 1+\frac{E_x-B_n}{4(B_n-\Delta)} \right),
\end{equation}
where $\sigma_0$ is the spin cut-off parameter at 
the neutron binding energy calculated according to \cite{Gil65}, and 
the pairing parameter $\Delta$ is the same as in Eq.\ (\ref{rhoEq}). 
This formula has the advantage that $\sigma(E_x)$ remains finite for all 
excitation energies and therefore one is not forced to make additional 
assumptions for $\sigma$ below $\Delta$.
A detailed description of the calculation of the integral in
Eq.\ (\ref{avgGamma}), including the necessary extrapolation of
experimental data to cover the energy region under consideration, 
is given in \cite{Voi01}.
The normalized experimental radiative strength functions for 
$^{170,171,172}$Yb are shown in Figure \ref{strall}.

It is assumed that the radiative strength is
dominated by dipole transitions. The Kadmenski{\u{\i}}-Markushev-Furman
(KMF) model is employed for the $E1$ strength. 
In the KMF model \cite{Kad83}, the Lorentzian GEDR is modified
in order to reproduce the non-zero limit of the GEDR for
$E_\gamma\rightarrow 0$ by means of a temperature-dependent
width of the GEDR. The $E1$ strength in the KMF model is given by
\begin{equation}
\label{KMF}
f_{E1}(E_\gamma)=\frac{1}{3\pi^2\hbar^2c^2}
\frac{0.7\sigma_{E1}\Gamma_{E1}^2(E_\gamma^2+4\pi^2T^2)}
{E_{E1}(E_\gamma^2-E_{E1}^2)^2},
\end{equation}
where $\sigma_{E1}$, $\Gamma_{E1}$, and $E_{E1}$ are the cross section,
width, and the centroid of the GEDR determined from photoabsorption
experiments. We adopt the KMF model with 
the temperature $T$ taken as a constant to be  consistent
with our  assumption  that the radiative strength function
is independent of  excitation energy.
The width of the GEDR is a sum of
energy and temperature dependent parts
\begin{equation}
\Gamma_{E1}(E_\gamma,T)=\frac{\Gamma_{E1}}{E_{E1}^2}
(E_\gamma^2+4\pi^2T^2).
\end{equation}
The giant dipole resonance is split into two parts for deformed nuclei.
Therefore, a sum of two strength functions each described by the above
equations is used. 

For the $M1$ radiation $f_{M1}$, the Lorentzian giant magnetic dipole
resonance (GMDR)
\begin{equation}
\label{GMDR}
f_{M1}(E_\gamma)=\frac{1}{3\pi^2\hbar^2c^2}
\frac{\sigma_{M1}E_{\gamma}\Gamma_{M1}^2}
{(E_\gamma^2-E_{M1}^2)^2+E_\gamma^2\Gamma_{M1}^2},
\end{equation}
is adopted.  This corresponds to  a spin-flip excitation.

A contribution from  $E2$ radiation is not included in 
Eq.\ (\ref{avgGamma}) because its strength is much smaller 
than the uncertainty due to the integration.
The Lorentzian $E2$ radiative strength  
\begin{equation}
\label{GEQR}
f_{E2}(E_\gamma) = \frac{1}{5 \pi^2 \hbar^2 c^2 E^2_{\gamma}}
\frac{\sigma_{E2} E_{\gamma}\Gamma_{E2}^2}{(E^2_{\gamma}-E_{E2}^2)^2+E^2_{\gamma}\Gamma_{E2}^2},
\end{equation}
is included in the summed radiative strength function for completeness.

For several rare-earth nuclei, an anomalous resonance
structure is observed in the radiative strength function
\cite{Voi01,Gut03b}. This resonance is observed in all rare-earth nuclei
that have been investigated by the Oslo method and is referred to as a 
pygmy resonance. In order to reproduce experimental results
where the pygmy resonance is observed, another Lorentzian centered
at $E_{\mathrm{py}}$ with width $\Gamma_{\mathrm{py}}$ and cross
section $\sigma_{\mathrm{py}}$ is used in addition to
the GEDR, GMDR, and the $E2$ resonance described above.
The total radiative strength function is composed of four parts
\begin{equation}
\label{SFfit}
f(E_\gamma)=\kappa (f^{I,II}_{E1}+f_{M1})+E^2_{\gamma}f_{E2}+f_{py},
\end{equation}
where $f^{I,II}_{E1}$ has the two components of the GEDR given
by the KMF model Eq.\ (\ref{KMF}), $f_{M1}$ and $f_{E2}$ are the 
giant magnetic dipole and electric quadrupole resonances 
given by Eqs.\ (\ref{GMDR}) and (\ref{GEQR}), respectively.
The parameters of these resonances are taken from \cite{RIPL}
and are listed
in Table \ref{tab2}. The parameters for the pygmy resonance $f_{py}$
and the overall multiplicative constant $\kappa$ were treated as 
fitting
parameters, as well as the $T$-parameter of the KMF model. 

The values obtained from the fit are listed in 
Table \ref{tab3}. 
The overall normalization factor $\kappa$ should be close to 1.
The deviation from 1 may be due to the normalization of the 
radiative strength function by the total radiative width 
or the approximation in the factor 0.7 in the KMF model. 
The energy and width of the pygmy resonance
in $^{171,172}$Yb from the different experiments agree well.

\begin{figure}
\includegraphics[totalheight=21cm,angle=0,bb=0 10 350 830]{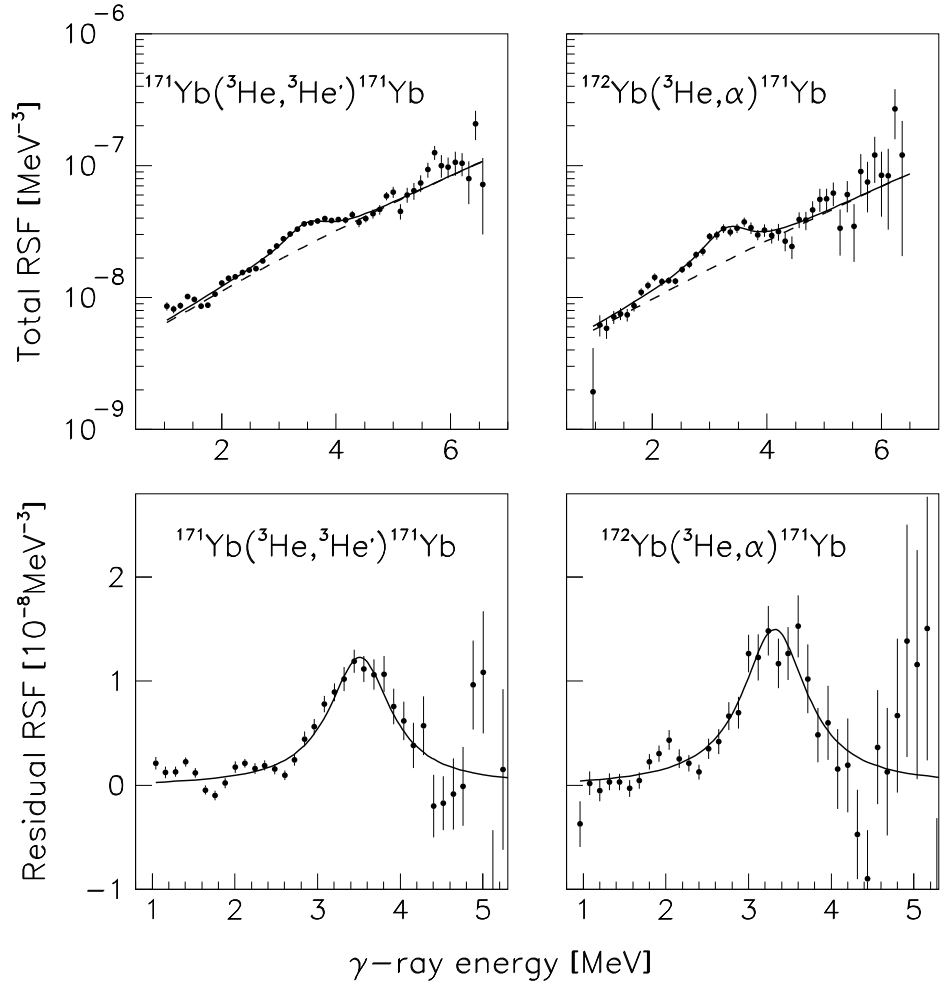}
\caption{A pygmy resonance in $^{171}$Yb observed in two reactions. 
The left two panels correspond to data obtained from the 
($^3$He,$^3$He$^{\prime}$) 
reaction. The two right panels show data from the ($^3$He,$\alpha$) reaction.
The solid line in the upper panels is a fit to data including all
contributions, the dashed lines are fits with  the contribution from
the pygmy resonance removed. 
The difference between the total $\gamma$-ray strength 
function data and the fit without the pygmy 
resonance (labeled as residual RSF)
is shown in the lower panels. The pygmy resonance is clearly identified.}
\label{pygmy}
\end{figure}

In Figure \ref{pygmy}, a fit to the experimental radiative strength 
function is shown. The upper panels contain the total radiative 
strength function (RSF) and the lower panels show the contribution 
from the pygmy resonance. After subtracting the fit
function without the pygmy resonances 
(dashed lines) from the data points of the upper
panel, the pygmy resonance is  clearly identified. The 
fit using only the pygmy resonances is shown as  solid lines in
the lower panels. 

\begin{figure}
\includegraphics[totalheight=21cm,angle=0,bb=0 10 350 830]{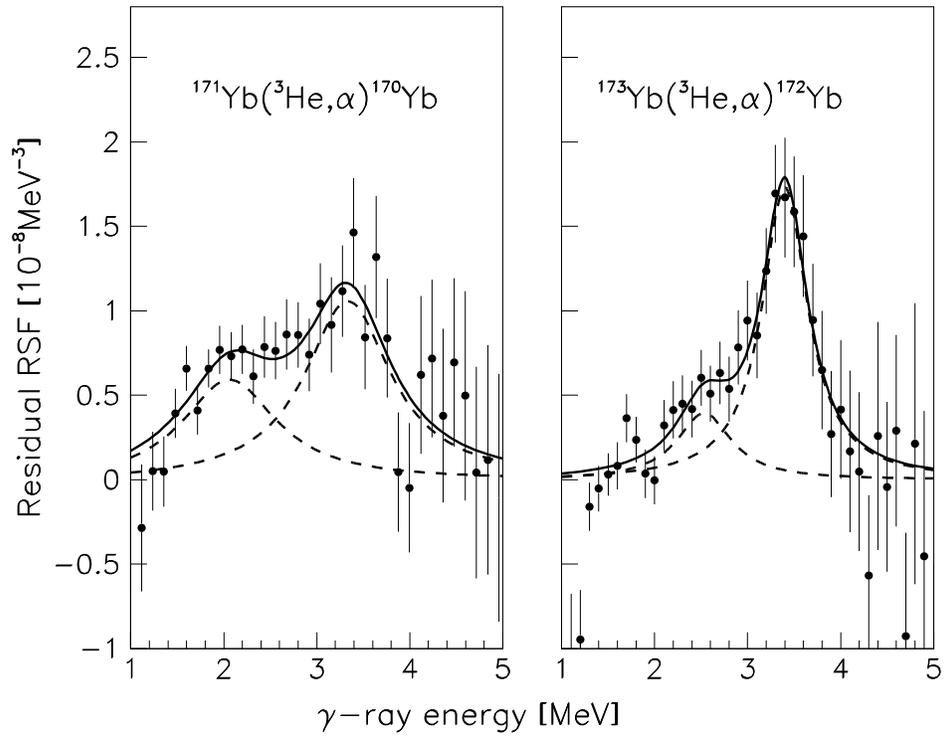}
\caption{ Two component fits of the pygmy resonances in $^{170,172}$Yb. 
The total fit (solid line) is described by the sum of two pygmy resonances
(dashed lines).}
\label{splitpy}
\end{figure}

The RSFs from the $(^3$He,$\alpha\gamma)$ reactions
are similar. The pygmy resonances seem to be split into
two components. The two-bump structure is so pronounced
in $^{170}$Yb  that a fit with $\kappa$ and $T$ as free
parameters failed. Therefore, the corresponding resonance parameters
could not be listed in Table \ref{tab3}. We fixed the $T$ parameter
from values for other Yb isotopes to be 0.34.
A similar splitting has been observed for Dy isotopes [23], 
although the fits with one component usually give satisfying results, e.g. for the
case of $^{172}$Yb, both the one- and two-component
fits give reasonable descriptions of the data. Examples of
two-component fits to the residual RSF (after subtracting
contributions from giant resonances) are shown in Figure \ref{splitpy},
the corresponding resonance parameters are given in Table \ref{tab4}.
By inspection of Figure \ref{pygmy}, a very weak structure at
$E_{\gamma} \sim 2.1$ MeV seems to be apparent in $^{171}$Yb as well.

In the case of $^{172}$Yb, the multipolarity of the pygmy
resonance has been established to be $M1$ \cite{Sch04ex}. The
resonance parameters are in reasonable agreement with
theory \cite{Lip83} and nuclear resonance fluorescence (NRF)
experiments \cite{Zil90} if we assume that for the scissors mode
in the quasicontinuum (above the pairing gap), the moment
of inertia is close to the rigid-body value and bare
$g$ factors have to be applied. NRF experiments on Dy
isotopes show that $M1$ excitations cluster around
$\sim 2.4$ and $\sim 3.0$~MeV \cite{Mar95}. In the present work,
the splitting into two components of the pygmy resonance
could be explained tentatively by the splitting in energy of
$\Delta K=\pm 1$ $M1$ $\gamma$ rays in the quasicontinuum.

\section{Conclusions}  

The level densities and radiative strength functions in $^{170}$Yb and
$^{171}$Yb are obtained from measured $\gamma$-ray spectra
following $^3$He induced reaction on $^{171}$Yb.
The deduced level densities extend structure data to excitation
energies above $\sim$ 2~MeV where the tabulated levels
are incomplete. The level densities and entropies for $^{170}$Yb and
$^{172}$Yb follow each other closely as a function of excitation energy.
The step structures in the level density 
indicate the breaking of the nucleon Cooper pair. 
The entropy carried by the valence neutron particle (or hole) 
in $^{171}$Yb is estimated to be $\Delta S=$ 2$k_B$ as expected.  
The radiative strength function in $^{171}$Yb exhibits
a resonance structure (pygmy resonance)
similar to that observed in a previous measurement.
The parameters for the pygmy resonance were obtained by fitting
the radiative strength function with common models, and
compared to values from the $^{172}$Yb($^3$He,$\alpha$)$^{171}$Yb
reaction. There is a good agreement between the two measurements.
The level density and strength function in $^{171}$Yb and $^{172}$Yb
using two different reactions give essentially the  same results 
leading to increased confidence to the applicability of 
statistical $\gamma$-ray spectroscopy.

\acknowledgements 

This research was sponsored by the National Nuclear Security Administration
under the Stewardship Science Academic Alliances program through DOE
Research Grant No. DE-FG03-03-NA00076. Support by U.S. Department of Energy 
Grant No. DE-FG02-97-ER41042 is acknowledged. 
Part of this work was performed under the auspices 
of the U.S. Department of Energy by the University of California, Lawrence 
Livermore National Laboratory under Contract No.\ W-7405-ENG-48.  
Financial support from the Norwegian Research Council (NFR) 
is acknowledged.

\end{multicols}
 
\begin{table}
\caption{Parameters used in the fits to the radiative strength functions.}
\begin{tabular}{l|lll}
                                & $^{170}$Yb & $^{171}$Yb & $^{172}$Yb\\ \hline
$E^I_{{\mathrm{E}}1}$, MeV      & 12.05      & 12.25      & 12.25 \\
$\sigma^I_{{\mathrm{E}}1}$, mb  & 239        & 239        & 239   \\
$\Gamma^I_{{\mathrm{E}}1}$, MeV & 2.78       & 2.6        & 2.6   \\ \hline
$E^{II}_{{\mathrm{E}}1}$, MeV     & 15.38      & 15.5       & 15.5 \\
$\sigma^{II}_{{\mathrm{E}}1}$, mb  & 302        & 302        & 302  \\
$\Gamma^{II}_{{\mathrm{E}}1}$, MeV & 4.64       & 4.8        & 4.8  \\ \hline
$E_{{\mathrm{M}}1}$, MeV        & 7.4        & 7.5        & 7.5  \\
$\sigma_{{\mathrm{M}}1}$, mb    & 1.30       & 1.50       & 1.76 \\
$\Gamma_{{\mathrm{M}}1}$, MeV   & 4          & 4          & 4 \\ \hline
$E_{{\mathrm{E}}2}$, MeV        & 11.37      & 11.35      & 11.33 \\
$\sigma_{{\mathrm{E}}2}$, mb    & 6.75       & 6.77       & 6.80 \\
$\Gamma_{{\mathrm{E}}2}$, MeV   & 4.07       & 4.06       & 4.05 \\ \hline
$\langle \Gamma_\gamma \rangle$, meV & 80(20)& 63(10) & 75(10)\\ 
\end{tabular}
\label{tab2}
\end{table}

\begin{table}
\caption{Fitted pygmy resonance parameters and normalization constants.}
\begin{tabular}{l|ccc|cc}
Reaction&$E_{\mathrm{py}}$&$\sigma_{\mathrm{py}}$&$\Gamma_{\mathrm{py}}$& T          & $\kappa$ \\
        &(MeV)            &(mb)                  & (MeV)                & (MeV) &
\\ \hline
&&&&\\
 
$^{171}$Yb($^3$He,$^3$He$^{\prime}$)$^{171}$Yb& 3.54(10) & 0.50(09)  & 0.91(18) & 0.31(2) & 1.26(06) \\
 
$^{172}$Yb($^3$He,$\alpha$)$^{171}$Yb         & 3.35(19) & 0.58(20) & 0.95(31) & 0.34(6) & 1.01(13) \\
 
$^{172}$Yb($^3$He,$^3$He$^{\prime}$)$^{172}$Yb& 3.28(18) & 0.48(12) & 1.36(41) & 0.33(4) & 1.65(11)  \\
 
$^{173}$Yb($^3$He,$\alpha$)$^{172}$Yb         & 3.38(27) & 0.58(29) & 0.99(55) & 0.37(5) & 1.85(17)\\
\end{tabular}
$^{a}$ The RSF of $^{170}$Yb could not be fitted by a single pygmy resonance, see text.
\label{tab3}
\end{table}

\begin{table}
\caption{Parameters for the two-component pygmy resonances}
\begin{tabular}{l|c|ccc}
Reaction& Component & $E_{\mathrm{py}}$&$\sigma_{\mathrm{py}}$&$\Gamma_{\mathrm{py}}$\\
        &           & (MeV)            &(mb)                  & (MeV)                \\\hline
        &           &                  &                      &                      \\
$^{171}$Yb($^3$He,$\alpha$)$^{170}$Yb         
        & I         & 2.15(21)         & 0.14(4)              & 1.23(53)             \\
$^{171}$Yb($^3$He,$\alpha$)$^{170}$Yb         
        & II        & 3.38(10)         & 0.41(7)              & 1.13(50)             \\ 
$^{173}$Yb($^3$He,$\alpha$)$^{172}$Yb         
        & I         & 2.56(16)         & 0.12(4)              & 0.72(34)             \\
$^{173}$Yb($^3$He,$\alpha$)$^{172}$Yb         
        & II        & 3.41(4)          & 0.68(9)              & 0.60(13)             \\
\end{tabular}
\label{tab4}
\end{table}

\end{document}